\documentclass[prl,twocolumn,a4paper,showpacs,showkeys]{revtex4}
\usepackage{amsmath,amsfonts}
\usepackage[latin1]{inputenc}
\usepackage{hyperref}
\usepackage{graphicx}

\renewcommand{\vec}[1]{\mathbf{#1}}
\newcommand{\avrg}[2][]{\langle #2 \rangle_{#1}}
\newcommand{\variance}[1]{\avrg{{#1}^2}}

\newcommand{\vacvar}{N_0}
\newcommand{\CoV}{\mathbf{K}}
\newcommand{\A}{\vec{A}}
\newcommand{\QA}{\vec{Q_A}}
\newcommand{\PA}{\vec{P_A}}
\newcommand{\QpA}{\vec{Q_{A'}}}
\newcommand{\PpA}{\vec{P_{A'}}}
\newcommand{\B}{\vec{B}}
\newcommand{\dB}{\delta B}

\newcommand{\QB}{\vec{Q_B}}
\newcommand{\PB}{\vec{P_B}}
\newcommand{\E}{E}

\newcommand{\ent}[2][]{H_{#1}(#2)}
\newcommand{\entz}{H_0}
\newcommand{\G}{\mathcal{G}}
\newcommand{\Proba}[1]{\mathcal{P}(#1)}
\newcommand{\R}{\mathbb{R}}

\begin{document}
\title{Continuous-variable quantum cryptography \\ 
is secure against non-gaussian attacks}
\date{\today}
%\author{Fr\'ed\'eric Grosshans and Nicolas J. Cerf}
\author{Fr\'ed\'eric \surname{Grosshans}} 
%\email[Permanent e-mail address: ]{frederic.grosshans@m4x.org}
%\altaffiliation[Current address: ]{Somewhere}
\author{Nicolas J. \surname{Cerf}}
%\email[E-mail address: ]{ncerf@ulb.ac.be} 
\affiliation{Quantum Information and Communication,
\'Ecole Polytechnique, CP 165, Universit\'e Libre de Bruxelles,
B-1050 Brussels, Belgium }


\begin{abstract}
  A general study of arbitrary finite-size coherent attacks 
  against continuous-variable quantum cryptographic schemes is presented. 
  It is shown that, if the size of the blocks that can be coherently attacked 
  by an eavesdropper is fixed and much smaller than the key size, then  
  the optimal attack for a given signal-to-noise ratio in the transmission line 
  is an individual gaussian attack. Consequently,
  non-gaussian coherent attacks do not need to be considered 
  in the security analysis of such quantum cryptosystems.
\end{abstract}
\maketitle

%\section{Introduction}

Continuous-variable quantum information theory has attracted a rapidly increasing interest over the past few years (see, e.g., \cite{BP}). In this context, several quantum key distribution (QKD) schemes based on the exchange 
of continuous key carriers have been proposed (see, e.g., \cite{hillery}).
In particular, several schemes based on the \emph{continuous} modulation 
of coherent or squeezed states of light supplemented with homodyne detection
have been shown to be particularly efficient for distributing secret keys
at high repetition rates \cite{CLVA,GG}. An experimental demonstration
of key distribution based on a gaussian modulation of coherent states was
recently provided in \cite{nature}.

In this Letter, we prove that given the estimated covariance matrix 
of Alice's and Bob's data, the optimal finite-size coherent attack
reduces to an individual gaussian attack characterized 
by this covariance matrix. This result fundamentally originates 
from the property that the distribution maximizing its Shannon entropy 
for a given variance is a gaussian distribution. This, combined with
an entropic uncertainty relation, implies that is is sufficient
to check the security of such cryptosystems against the restricted 
class of gaussian attacks.
% (it is unnecessary to consider non-gaussian attacks).
In other words, the best strategy for Eve is to apply 
sequentially, on each key element, a gaussian cloning machine \cite{CIR} or an entangling gaussian cloning machine \cite{GG-Proc} depending on the exact protocol used. Another consequence is that, in order to maximize the
resulting secret key rate via the gaussian channel induced by Eve's attack,
Alice should modulate her data with a gaussian distribution.

The security proof presented here is valid for all continuous-variable
QKD schemes where Alice and Bob monitor the transmission line via the 
second-order moments of their data, which includes all the protocols
considered in our previous papers \cite{CLVA,GG,nature}. Note, however,
that this excludes the alternative protocol based on postselection
as presented in \cite{silberhorn}. Our proof covers all possible 
(including coherent) attacks that an eavesdropper may apply 
on \emph{finite-size} blocks of key elements.
The block size may be arbitrary, but it must be much smaller than the key size,
so that the key is made out of a large number of independent blocks
and statistical arguments therefore warrant the use of information theory
in the proof. The unconditional security of squeezed-state QKD against coherent attacks is currently proven if the squeezing exceeds
some threshold \cite{gottesman}, while such a proof for coherent-state QKD
is the topic of a separate study \cite{IVAC}.

\paragraph{Squeezed state protocols.}
\label{sec:EPR}

Let us first investigate the security of gaussian-modulated
squeezed-state protocols \cite{CLVA}. Alice chooses a quadrature ($q$ or $p$)
at random and sends Bob a displaced squeezed state, where the squeezing and displacement are applied on the chosen quadrature while the value of the
displacement is gaussian distributed. After transmission via the quantum 
channel, which may be controlled by Eve, Bob then measures $q$ or $p$ at random.
After disclosing the quadrature they used, Alice and Bob discard their data 
when the quadratures differ, while the rest is used to make
a secret key \cite{slice, supplinfo}. We will in fact consider equivalent
entanglement-based protocols \cite{GG}, where Alice prepares a two-mode
vacuum squeezed state, measures a quadrature of one of the beams and sends 
the other beam to Bob.
Alice and Bob iterate these actions $n$ times, while we assume that
Eve is able to apply some arbitrary joint operation on this block
of $n$ pulses. 
In order to acquire accurate statistics,
Alice and Bob repeat this protocol $L$ times (with $L\gg 1$), that is,
they exchange $L$ blocks of $n$ pulses in total. 
%Actually, in order to
%obtain a good statistical estimate of the covariance matrix, Alice and Bob
%will also have to sacrifice a small fraction of these data.
In our security analysis below, we will apply information theory 
at the level of blocks, which is justified since $L\gg 1$. 
%This is also 
%why we only cover \emph{finite-size} collective attacks: 
%$n$ can be as big as one wants, but one needs to exchange
%$L\times n\gg n$ pulses for our proof to guarantee the security of
%the protocol.

\begin{figure}[tb]
\includegraphics{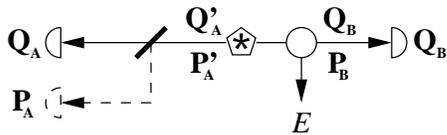}
\caption{\textbf{Equivalent entanglement-based QKD protocol.} 
The twin beams of an EPR-source (*) are sent to homodyne 
detectors at Alice's (left) and Bob's (right) side. In the analogue
of the squeezed-state protocol, the beam-splitter and the dashed lines are omitted, so Alice only measures one quadrature ($\QpA=\QA$). In the
analogue of the coherent-state protocol, the beam-splitter is used by Alice
to measure $\QpA$ and $\PpA$ simultaneously \cite{virtual}.}
\label{fig}
\end{figure}

We model Eve's attack by considering that Alice, Bob, and Eve share a
pure tripartite entangled state (see Fig.~\ref{fig}). 
Alice's (resp. Bob's) part of the state 
is a set of $n$ modes, denoted by $\vec{A}$ (resp. $\vec{B}$). The unknown physical system kept by Eve is denoted by $\E$. The joint
state is pure since we must assume that Eve is able to control the
environment, thereby to purify the state.
We also suppose that 
%Alice chooses at random to
%use the quadrature $q$ or $p$ for each pulse, and that 
Bob always measures the same quadrature $Q$ as Alice ($Q=q$ or $p$). This requires the availability of a quantum memory (Bob delays his measurement 
until Alice discloses the quadrature she used).
In a more realistic scheme where Alice and Bob independently choose their quadrature $q$ or $p$ at random, they agree only half of the time, 
which simply leads to a factor 1/2 
in the information rates computed below.

\paragraph{Information Rates.}

The mutual information between Alice's and Bob's data is
\begin{align}
  I(\B;\A)=I(\QB;\QA)
          =\ent{\QB}-\ent{\QB|\QA},
\end{align}
where $\QA$ (resp. $\QB$) is the random vector of Alice's
(resp. Bob's) measured quadratures on a block of $n$ pulses, 
while $\ent{\cdot}$ [resp. $\ent{\cdot|\cdot}$] denotes the Shannon entropy 
(resp. conditional entropy) for continuous random variables.
We focus our attention on reverse reconciliation protocols \cite{nature,GG-Proc}, in which Bob's data are used to make the key
instead of Alice's data (direct reconciliation). Then,
Eve tries to get the maximum information on Bob's measurement
outcomes $\QB$ through a measurement of her ancilla $\E$
(we denote Eve's ancilla and her measurement outcomes
by the same symbol $\E$). Eve's information is
\begin{align}
  I(\B;E)=I(\QB;E)
        =\ent{\QB}-\ent{\QB|\E}.
\end{align}
The secret key rate Alice and Bob are guaranteed to be able to 
distill by reverse reconciliation is \cite{csiszar,maurer}
\begin{align}
  \Delta I&=I(\B;\A)-I(\B;E)\nonumber\\
          &=\ent{\QB|E}-\ent{\QB|\QA}.
  \label{DeltaI}
\end{align}
Alice and Bob can, in principle, estimate $\ent{\QB|\QA}$ with arbitrary
precision since they have access to $L$ joint realizations 
of the random vectors $\QA$ and $\QB$. 
To lower bound Eve's uncertainty on the key $\ent{\QB|E}$, 
they can use the entropic uncertainty relation that applies to
the two sets of conjugate quadratures $\QB$ and $\PB$ \cite{bial,beckner}.
Indeed, we know that by measuring their systems, Alice and Eve 
project Eve's system onto a pure state since the three of them 
share a joint pure state. 
Thus, conditionally on Alice's and Eve's measurements $\PA$ and $E$, the pure
state held by Bob must satisfy the entropic inequality
\begin{equation}
  \label{eq:Shannonberg}
  \ent{\QB|E}+\ent{\PB|\PA}\ge2n\entz,
\end{equation}
where $\entz$ is the entropy of a quadrature of the vacuum
state for an harmonic oscillator. 
This inequality then allows us to lower bound the accessible secret key rate
regardless the action of Eve, namely
\begin{equation}
  \Delta I\ge\Delta I_{\min}=2n\entz-\ent{\QB|\QA}-\ent{\PB|\PA}.
\end{equation}

It is worth stressing that the random vectors $\PA$ and $\PB$
%. As already mentioned, $\QA$ and $\QB$ denote the quadratures
%(either $p$ or $q$ ) that are measured by Alice and Bob on a block of
%$n$ pulses in order to make the key. In contrast,
%$\PA$ and $\PB$ 
denote the quadratures that \emph{could have} been measured (the measured
quadratures are $\QA$ and $\QB$).
These quadratures are, of course, not directly accessible, but we only
need their statistical distribution here in order to upper bound Eve's
information. This distribution can be estimated
from the other pulses for which the measured quadrature 
is the same as $\PA$ and $\PB$.
%but their 
%statistical distribution can be estimated from the other half of the key elements for which the
%measured quadratures $\QA$ and $\QB$ are the same physical 
%quadrature (i. e., $p$ or $q$) as $\PA$ and $\PB$.
For simplicity, we assume that the two physical quadratures 
$q$ and $p$ are both chosen with probability 1/2. This implies that
$\QA$ and $\PA$ play fully identical roles so they can be treated completely symmetrically (the same is true for $\QB$ and $\PB$). 
We insist on that this symmetry is not a limitation on Eve's possible 
actions. Even if Eve has a quantum memory and acts differently on the physical
quadratures $q$ and $p$ (after the selected quadrature is disclosed), each
of them has an equal probability to be a measured
($\QA$ and $\QB$) or an unmeasured ($\PA$ and $\PB$) quadrature.
Since Eve has no way of guessing which physical quadrature is used,
this symmetry imposes 
$\ent{\QB|\QA}= \ent{\PB|\PA}= \ent{\B|\A}$,
where we now use $\A$ and $\B$ as a shorthand notation for $\QA$ and $\QB$
(or $\PA$ and $\PB$). Therefore
\begin{equation}
  \label{eq:DIminSq}
  \Delta I_{\min}=2(n\entz-\ent{\B|\A}).
\end{equation}
Since Alice and Bob can evaluate $\ent{\B|\A}$ by statistical sampling,
they get an estimate of $\Delta I_{\min}$ and can use relevant algorithms 
to extract a secret key with at least this rate \cite{slice,supplinfo}. 

\paragraph{Individual attacks are optimal.}
We first prove that Alice and Bob can 
lower bound $\Delta I_{\min}$ simply by assuming that Eve 
performs an individual attack.
Let $A_i$ (resp. $B_i$) be the $i$th component of the random vector $\A$
(resp. $\B$). The subadditivity of Shannon entropy implies that
\begin{equation}
  \ent{\B|\A}\le\sum_i\ent{B_i|\A},
\end{equation}
%Using the fact that conditioning can only decrease entropy, one has 
while each term of the summation can be bounded by use of the strong subadditivity of the entropy, namely
\begin{equation}
  \ent{B_i|\A}=\ent{B_i|A_1,\dots,A_n}\le \ent{B_i|A_i},
\end{equation}
so that
\begin{equation}  \label{eq:individual}
  \ent{\B|\A}\le\sum_i\ent{B_i|A_i}.
\end{equation}
We now consider the \emph{average} joint distribution of Alice's and Bob's
measurement outcomes (averaged over the block of size $n$).
Suppose that $A$ and $B$ are distributed according 
to a mixture of the $A_i$'s and $B_i$'s, with the index $i$ being 
randomly drawn from a uniform distribution, that is
\begin{equation}
  \Proba{A=a,B=b}=\frac1n\sum_i\Proba{A_i=a,B_i=b}, \quad \forall a,b.
\end{equation}
%In other words, the index is considered like a random variable $I$,
%and it's value $i$ is forgotten. One has then $A=A_i$ when $I=i$. 
Then, the strong subadditivity of entropies implies that
\begin{equation}  
\ent{B_i|A_i}=\ent{B|A,i}\le\ent{B|A},
\end{equation}
%and one has
%\begin{equation}
%  \sum_i\ent{B_i|A_i}\le n\ent{B|A},
%\end{equation}
so that Eq.~(\ref{eq:individual}) transforms into
\begin{equation}
  \ent{\B|\A}\le n\ent{B|A}.
\end{equation}
Finally, using Eq.~\eqref{eq:DIminSq}, one gets
\begin{equation}
  \label{eq:DIminSQInd}
  \Delta I_{\min}\ge 2n(\entz-\ent{B|A}).
\end{equation}
This means that, to be safe against finite-size coherent attacks,
Alice and Bob only need to evaluate $\ent{B|A}$, 
a conditional entropy for a distribution in $\R^2$, instead of
$\ent{\B|\A}$, a conditional entropy for a distribution in $\R^{2n}$.

To better understand this conclusion, assume that Eve
applies a coherent attack which induces correlations between the various
components of $\A$ and $\B$ inside each block. These correlations 
force Eve to induce a kind of structure in Alice's and Bob's data, 
which would not be present for individual attacks, so Eve is actually
limiting herself. Overlooking these correlations and considering 
individual attacks only may be suboptimal for Alice and Bob 
when estimating $\Delta I_{\min}$, but it guarantees 
they are on the safe side. 
%Since the evaluation of $\ent{B|A}$ can be done by measuring the
%noise variance only, as we shall see, and since the
%structure of naturally occurring noise often mimics individual
%attacks, these will usually limit themselves to individual attacks.

\paragraph{Gaussian attacks are optimal.}

Now, we prove that $\ent{B|A}$ can be upper bounded simply 
by measuring the covariance matrix $\CoV$ of variables $A$ and $B$, 
\begin{equation}
  \CoV=
  \begin{bmatrix}
    \variance{A} & \avrg{AB}\\
    \avrg{AB} & \variance{B}
  \end{bmatrix},
\end{equation}
which is much easier than estimating $\ent{B|A}$.
% as the latter 
% is a non-linear function.
%Since the entropy is a non-linear function, evaluating the entropy of a
%distribution is more difficult than evaluating its covariance matrix. 
To simplify the notations, we will assume that
$\avrg{A}=\avrg{B}=0$ (in practice, this should be checked and
possibly corrected by applying the adequate shift).
 %\begin{equation}
%  \ent[\G]{B|A}=\tfrac12\log2\pi\e
%%  +\tfrac12\log\left(\variance{B}-\frac{\avrg{AB}^2}{\variance{A}} \right)
%   +\tfrac12\log\frac{\det{\CoV}}{\variance{A}},
%\end{equation}
%where $\det{\CoV}$ is the determinant of the covariance matrix
%$\CoV$. 
For a given $\CoV$, if Alice knows $A$, her linear estimate
of $B$ that minimizes the error variance is given by
$\tfrac{\avrg{AB}}{\variance{A}}A$. Denoting by $\dB$
the error of this best linear estimate,
\begin{equation}
  \dB=B-\frac{\avrg{AB}}{\variance{A}}A
\end{equation}
we have
\begin{equation}
  \label{eq:ng2g1}
  \ent{B|A}=\ent{\dB|A}\le\ent{\dB}.
\end{equation}
where we have used the translation invariance and the subadditivity of 
Shannon entropy.
%Furthermore, the strong subadditivity of the entropy gives us
%\begin{equation}
%  \label{eq:ng2g2}
%  \ent{\dB|A}\le\ent{\dB}.
%\end{equation}
Since the gaussian distribution has the maximum entropy for a given variance,
one has
\begin{equation}
  \ent{\dB}\le\ent[\G]{\dB},% =\tfrac12\log2\pi\e+\tfrac12\log\variance{\dB}.
\end{equation}
where $\ent[\G]{\dB}$ is the entropy of a gaussian distribution having
the variance $\variance{\dB}=\avrg{B^2}-\avrg{AB}^2/\avrg{A^2}$.
%Computing $\variance{\dB}$, one finds 
%\begin{equation}
%  \avrg{\dB^2}=\avrg{B^2}-\frac{\avrg{AB}^2}{\avrg{A^2}}
%      =\frac{\det{\CoV}}{\variance{A}}, 
%\end{equation}
%hence
In the case where $A$ and $B$ are drawn from
an equivalent bivariate Gaussian distribution
with the same covariance matrix $\CoV$ as the observed distribution,
we note that $\dB$ and $A$ become uncorrelated, so that
\begin{equation}
  \label{eq:ng2g3}
  \ent[\G]{\dB}=\ent[\G]{\dB|A}
\end{equation}
%Since the entropy is invariant by translation, one has, 
%\begin{equation}
%  \label{eq:ng2g4}
%  \ent[\G]{\dB|A}=\ent[\G]{B|A},
%\end{equation}
%which is very similar to equation \eqref{eq:ng2g1}
Chaining Eqs.~\eqref{eq:ng2g1} to \eqref{eq:ng2g3} 
and using the translation invariance of entropy, one obtains
\begin{equation}
  \ent{B|A}\le\ent[\G]{B|A},
\end{equation}
%which proves that $\ent{B|A}$ can indeed be upper bounded by
%considering the equivalent gaussian distribution.
%for a given covariance matrix $\CoV$.
which, combined with Eq.~\eqref{eq:DIminSQInd}, yields 
\begin{equation}
  \label{eq:DIminSQGbis}
  \Delta I_{\min}\ge2n(\entz-\ent[\G]{B|A}).
\end{equation}
Finally, the conditional entropy $\ent[\G]{B|A}$ of a bivariate gaussian distribution being
a simple function of $\CoV$, one obtain the
central result of this paper,
\begin{equation}
  \label{eq:DIminSQG}
  \Delta I_{\min}\ge 
   n \log\frac{\vacvar}{\variance{\delta B}},
\end{equation}
%\begin{equation}
%  \label{eq:DIminSQG}
%  \Delta I_{\min}\ge n\log\frac{\vacvar\variance{A}}{\det{\CoV}}
%    =n\log\frac{\vacvar}{\variance{B}-\frac{\variance{AB}}{\variance{A}}},
%\end{equation}
where $\vacvar$ represents the vacuum variance.
This expression coincides with the one found when limiting Eve to
gaussian individual attacks \cite{GG,nature}. Therefore, 
the optimal attack given the observed covariance matrix $\CoV$
is a gaussian individual attack as described in \cite{nature,GG-Proc,virtual}.

The optimality of gaussian attacks can be interpreted almost alike
the optimality of individual attacks : 
since the gaussian distribution has the maximal entropy, 
non-gaussian attacks are more structured than
gaussian ones for a same added noise variance, 
so Eve is more restricted. Therefore, if Alice and Bob
only monitor the covariance matrix $\CoV$, they can safely
assume that Eve uses gaussian attacks. If Eve indeed applies a gaussian
attack, the best Alice and Bob can do is to use independent 
and gaussian-distributed key elements, which saturates all the 
involved inequalities, so that $\Delta I_{\min}$ is the highest.
This justifies \emph{a posteriori} the choice of gaussian-modulated
QKD protocols in \cite{CLVA,GG,nature}.

\paragraph{Coherent state protocols.}

We now extend the proof to QKD protocols based on 
gaussian-modulated coherent states \cite{GG,nature,GG-Proc}. 
We again exploit the property that these protocols are
equivalent to some entanglement-based protocols where
Alice jointly measures $q$ and $p$ on her entangled beam 
while sending the other one to Bob \cite{virtual}.
The central point is that this ``virtual entanglement,'' which 
may have existed between Alice and Bob, must be taken into
account when bounding Eve's information even if the actual protocol
makes no use of entanglement.
 %one must take into account 
%the ``virtual'' entanglement of a protocol that is fully equivalent 
%to the coherent-state protocol, as explained in  
%The above considerations about squeezing seem to prevent the
%extension of the previous results to coherent state quantum
%cryptography.
%However, similarly to \cite{nature, GG-Proc,virtual}, 
%one can consider the squeezing that \emph{could have been present}. 
%However, as shown in \cite{virtual}, a coherent state 
%quantum key distribution protocol is
We will denote by $\QpA$
and $\PpA$ the vectors of the two quadratures of the $n$ beams kept by Alice and $\QB$ the vector of the $n$ quadratures measured by
Bob (see Fig.~\ref{fig}). 
The difference with the previous scheme is that Alice attempts to measure simultaneously $\QpA$ and $\PpA$ through a 50:50 beam-splitter followed by 
two homodyne detectors. The measurement outcomes $\QA$ and $\PA$ suffer from added noise, while Alice never has access to the actual values 
$\QpA$ and $\PpA$. The expression of $\ent{\QB|\QA}$ only depends on measured quantities so it can be statistically estimated as before,
%\begin{equation}
%  \Delta I=\ent{\QB|E}-\ent{\QB|\QA}.
%\end{equation}
but the entropic uncertainty relation used to bound $\ent{\QB|E}$ 
now involves the physical beam on Alice's side, so one has
\begin{equation}
  \ent{\QB|E}+\ent{\PB|\PpA}\ge2n\entz
\end{equation}
Thus, the same reasoning as before now leads to
\begin{equation}
  \Delta I_{\min}\ge n(2\entz-\ent[\G]{B|A}-\ent[\G]{B|A'}),
\end{equation}
where $\ent[\G]{B|A'}$ is the conditional entropy of a gaussian distribution
having the same covariance matrix $\CoV'$ than $A'$ and $B$, which is
%$\ent[\G]{B|A'}$ is a function of $\CoV'$, which can be computed from $\CoV$:
\begin{equation}
  \CoV'=
  \begin{bmatrix}
    \variance{A'} & \avrg{A'B}\\\avrg{A'B}&\variance{B}
  \end{bmatrix}
   =
  \begin{bmatrix}
    2(\variance{A}-\vacvar) & {\sqrt2}\avrg{AB}\\
     {\sqrt2}\avrg{AB}&\variance{B}
  \end{bmatrix}
\end{equation}
One has therefore
\begin{equation}
  \label{eq:DIminCG}
    \Delta I_{\min}\ge
     n\log
     \frac{\vacvar}
	{\sqrt{\bigl(\variance{B}-\frac{\avrg{AB}^2}{\variance{A}}\bigr)	      \bigl(\variance{B}-\frac{\avrg{AB}^2}{\variance{A'}-\vacvar}\bigr)}},
\end{equation}
which is exactly the same expression as in our previous papers
\cite{nature, GG-Proc,virtual}, where the only considered attacks
are gaussian individual attacks.

\paragraph{Discussion.}

We extended to finite-size non-gaussian attacks the validity of the previous security proofs for continuous-variable QKD schemes when Eve's intervention
is bounded via the measured added noise variance in the channel. Our proof 
focuses on the schemes based on reverse reconciliation since these
are known to tolerate larger losses than the direct reconciliation-based protocols in the case of gaussian individual attacks. Adapting the proof
to direct-reconciliation \cite{CLVA,GG} or even other \cite{silberhorn}
protocols will treated elsewhere.
%but it should be easily adapted to the case of direct reconciliation.
In the proof, we assume the protocol is ideal, that is,
a perfect one-way reconciliation algorithm is available. 
However, realistic reconciliation protocols are imperfect \cite{supplinfo}:
the number of correlated bits that can be extracted from Alice's
and Bob's data never attains Shannon's limit $I(B;A)$ and may
become low if Eve's attack
has an unexpected shape, the reconciliation protocol being adapted to
a specific noise structure.
%Furthermore, the value of $I(B;A)$ can be further decreased if Eve's attack
%has an unexpected shape, the reconciliation protocol being adapted to
%a specific noise structure.
%the sliced reconciliation protocol we used in
%experimental continuous variable cryptography \cite{slice,nature} 
%is neither one-way nor perfect. It is based a slicing procedure
%supplemented with a reconciliation protocol named \emph{Cascade}, which is heavily interactive so it cannot be considered one-way. This
%difficulty was previously overcome by a numerical computation of the
%information leaked in the case of the optimal gaussian attack\cite{supplinfo},
%but this approach seems difficult to generalize to the non-gaussian 
%coherent case. 
%However, recent work [Cardinal-Van Assche
%03] gives hope to have efficient one-way reconciliation protocols in
%the near future.
Nevertheless, the security proof can be easily
extended to this situation since Alice and Bob can always
compute the effective value of their shared information $I_{\rm eff}$ by comparing subsets of their data. Then, 
using $I(B;E)\le n(\entz-\ent[G]{B|A'})$ as before, one
obtains $\Delta I_{\min}\ge I_{\rm eff}- I(B;E)$.

Finally, we have shown that there is a fundamental link 
between security and ``entropic squeezing'': the security is guaranteed 
($\Delta I_{\min}>0$) if the conditional entropy $\ent{B|A}$ 
is below the quantum limit $\entz$ [Eq.~\eqref{eq:DIminSQInd}]. 
%This is more general that the link between security and
%squeezing that we used in our previous papers. Noting that
%the denominator of Eq.~\eqref{eq:DIminSQG} is 
In the gaussian case, this simplifies to condition 
%if $B$ is squeezed conditionally on $A$ 
$\sigma^2(B|A) < \vacvar$ [Eq.~\eqref{eq:DIminSQG}], where 
$\sigma^2(B|A) = \variance{\delta B}$
denotes the conditional variance of $B$ knowing $A$,
as suggested in \cite{PRA}.
%In contrast, denoting the conditional
%variance of $B$ knowing $A$ as $\sigma^2(B|A) = \variance{\delta B}$, 
%Eq.~\eqref{eq:DIminSQG} implies that the security is guaranteed
%if $B$ is squeezed conditionally in $A$ 
%(i.e., $\sigma^2(B|A) < \vacvar$).
The latter condition is, however, over-pessimistic
if Eve uses a non-gaussian attack, since
$\sigma^2(B|A)$ might exceed $N_0$, destroying the conditional squeezing, while keeping $\ent{B|A}$ low enough to ensure security. 
If Alice and Bob only monitor the covariance matrix $\CoV$, this attack is
non-optimal since the worst-case gaussian attack would
maximize $\ent{B|A}$ and thereby minimize $\Delta I$ for
a given $\sigma^2(B|A)$. In conclusion, the security can be warranted
by requiring conditional squeezing, which is more stringent
than entropic squeezing but much easier to assess.

%Alice and Bob are sure to be on the safe side. 

%\comment[FG]{Sinon, seulement attaques individuelles (nous +
%Erlangen). Adaptation a  leurs postselection based schemes encore a 
%faire.}

%\comment[FG]{Parler de Gottesmann-Preskill, puis Iblisdir, GVA,
%  Cerf. Attaques plus generale, mais: squeezing necessaire pour le
%  premier, et portee encore limitee pour les deux.}

%\comment[FG]{Adaptation aux protocoles directs triviale ds le premier
%  cas. A faire dans le cas des protocoles a etats coherents.}

%\section*{Acknowledgments}
\begin{acknowledgments}
  We are grateful to Philippe Grangier for stimulating this work, 
  and to Patrick Navez and Gilles Van Assche for useful discussions. 
  FG acknowledges support from the Belgian National
  Fund for Scientific Research. NJC acknowledges financial support from the
  Communauté Française de Belgique under grant ARC 00/05-251, from the IUAP
  programme of the Belgian governement under grant V-18, and from the EU under
  project RESQ (IST-2001-35759).
\end{acknowledgments}

\end{document}